\documentclass[%
 reprint,
 amsmath,amssymb,
 aps,
prl,
]{revtex4-2}

\usepackage[utf8]{inputenc}
\usepackage{graphicx}
\usepackage{dcolumn}
\usepackage{bm}
\usepackage{amsfonts}
\usepackage{color}

\begin{document}

\preprint{APS/123-QED}

\title{Anomalous Resistivity at Weak Coupling}

\author{Thomas G. Kiely}
 \altaffiliation{tgk37@cornell.edu}
\author{Erich J. Mueller}
 \altaffiliation{em256@cornell.edu}
\affiliation{Laboratory of Atomic and Solid State Physics, Cornell University, Ithaca, NY 14853}

\date{\today}

\begin{abstract}
Recent cold atom experiments have observed bad and strange metal behaviors in strongly-interacting Fermi-Hubbard systems. Motivated by these results, we calculate the thermoelectric transport properties of a 2D Fermi-Hubbard system in the weak coupling limit using quantum kinetic theory. We find that many features attributed to strong correlations are also found at weak coupling.
In particular, for temperatures $T\gtrsim t$ the electrical resistivity is nearly linear in temperature 
despite the fact that the quasiparticle scattering rate 
 is non-linear and changes
by nearly an order of magnitude.
We argue that this asymptotic behavior is a general feature of systems with a finite spectral width, which implies that there is no MIR bound on the resistivity in single-band models.
Due to nesting, the $T$-linear resistivity persists down to $T=0$ at half filling.
Our work sheds light on the transport regime in ultracold atom experiments, which can differ substantially from that of condensed matter systems.
Disentangling these band-structure effects from the physics of strong correlations is a major challenge 
for future experiments.
\end{abstract}

\maketitle

Anomalous transport properties in the normal state of various strongly-correlated materials~\cite{Hussey,ruthenate,pnictide,heavyFermion,univPlanck,scattRt} have sparked considerable interest in recent years. 
The terms ``strange metal" and ``bad metal" have come to denote two significant deviations from standard Fermi liquid behavior: a DC resistivity which is $T$-linear down to low temperatures and an inferred mean-free path which becomes shorter than the lattice spacing at high temperatures, respectively. By contrast, the DC resistivity of a Fermi liquid is expected to vanish as $T^2$ at low temperatures and saturate at an upper bound, known as the Mott-Ioffe-Regel (MIR) limit~\cite{coleman,mott,ioffe}. The stark deviation of these behaviors from Fermi liquid theory has inspired a wealth of sophisticated theoretical and numerical approaches~\cite{cuprateNS,critpt,holographic,highTPerspective,hartnollBadMetal,numericalHubbardVertex,Huang,slopeInvariantTLinear}, many of which rethink the notion of Fermi liquid theory altogether.

Recent developments in cold atoms~\cite{expReview1,expReview2} have lead to exciting experiments that shed a unique light on this transport problem~\cite{Brown379,demarco,thywissen}. 
The neutral atoms are assigned a fictitious charge $e$, forces are interpreted as electric fields $E=F/e$, and the language of electronic systems is used in order to connect to solid-state problems.
These cold atom experiments operate in a novel transport regime in which an ``electronic" current can only dissipate via electron-electron interactions. 
Studying transport in this regime will help to disentangle the role of interactions, phonons, and impurities in real materials. Furthermore, cold atom experiments are poised to study fermionic systems with a wide variety of interaction strengths, particle densities, and temperatures. Already, the Bakr group has measured a $T$-linear DC resistivity in the 2D Fermi-Hubbard model that smoothly violates the MIR limit~\cite{Brown379}. Their experiment further illuminates the problem by studying the Nernst-Einstein decomposition of the conductivity into the product of the diffusion constant and the charge compressibility, $\sigma=D\chi$. Interestingly, they find that the diffusion constant and the quasiparticle scattering rate saturate at a high-temperature asymptotic value. This saturation is consistent with the notion of an MIR bound, even though the electrical resistivity appears to be unbounded. Furthermore, they find that the diffusion constant and the charge compressibility exhibit non-trivial temperature dependences at intermediate temperatures which ``conspire" in such a way that the DC resistivity, $\rho=1/\sigma$, is $T$-linear with a constant slope. 

In this Letter, we study transport in the 2D Fermi-Hubbard model in the weak-coupling regime where Fermi liquid theory is applicable. We find a number of striking similarities to the Bakr experiment including a high-temperature $T$-linear DC resistivity that persists to intermediate-to-low temperatures. We explain the high temperature behavior in terms of a diverging effective mass.  Consistent with the MIR limit, however, the scattering rate remains finite. As observed in Ref.~\cite{Brown379}, the $T$-linear resistivity at intermediate temperatures coincides with  non-trivial temperature dependence of the charge compressibility and diffusion constant.  At half-filling, the $\rho\propto T$ behavior continues down to $T=0$, where it can be attributed to the nested bandstructure.
These features of the weakly-coupled system must be taken into account when looking for strong-correlation physics in cold atom experiments.


The Fermi-Hubbard model is described by the Hamiltonian
\begin{multline}
    \mathcal{H}=-t\sum_{\langle ij\rangle,\sigma}(c^\dagger_{i\sigma}c_{j\sigma}+h.c.)
    +U\sum_ic^\dagger_{i\uparrow}c^\dagger_{i\downarrow}c_{i\downarrow}c_{i\uparrow}
\end{multline}
where $c^\dagger_{i\sigma}$ is the fermionic creation operator on site $i$ with spin $\sigma$. It describes particles that hop between sites on a lattice with a tunneling matrix element $t$ and onsite repulsion $U$. This is a natural model for fermionic atoms in optical lattice, where $t$ is set by the lattice depth and $U$ can be tuned by a Feshbach resonance~\cite{expReview1,expReview2}.  It is also hypothesized to be a minimal model for understanding a variety of strongly-correlated condensed matter systems~\cite{ANDERSON1196}. In the absence of interactions, the kinetic energy is diagonalized in momentum space with a dispersion $\epsilon_k=-2t\cos{k_x}-2t\cos{k_y}$. Thus, the non-interacting bandwidth is $W=8t$.

While this simple model has a rich phase diagram at strong coupling~\cite{arovas2021hubbard,computationalPerspective}, it can be described in terms of Fermi liquid theory when $U/t\ll 1$. We model transport in this regime using quantum kinetic theory, following the methods of Refs.~\cite{Ziman,Ziman2} that are described in detail in a companion paper~\cite{us}. We study the behavior of the steady-state distribution function $f_k(r)$, which corresponds to the particle number density in a differential volume of phase space $d^2k~d^2r/(2\pi)^2$. The distribution function is related to the real-space number density via $n(r)=2\int\frac{d^2k}{(2\pi)^2}f_k(r)$ where the factor of $2$ is for spin. It evolves in time according to a Boltzmann equation,
\begin{equation}
    \partial_tf_k+\nabla_rf_k\cdot {\bf v}_k + e{\bf E}\cdot\nabla_kf_k=I_k[f],
    \label{eq:boltzeq}
\end{equation}
where the collision integral, $I_k[f]$, accounts for the effects of quasiparticle scattering. Although in cold atom experiments the particles are neutral, we use the language of electronic systems, formally assigning them a charge $e$.  As already explained, the quantity $e{\bf E}$ is simply a force, for example from optical or magnetic fields.

We expand about the equilibrium distribution function
$f^0_k=\big(1+e^{\beta(\epsilon_k-\mu)}\big)^{-1}$, parameterizing the deviation in 
terms of a set of 
trial functions. The coefficients of these trial functions are then determined using Onsager's condition that the steady-state distribution maximize entropy production~\cite{Ziman,Ziman2,onsager1,onsager2}. Knowledge of these coefficients allows us to determine the full thermoelectric matrix at a given temperature and particle density.

\begin{figure}
    \centering
    \includegraphics[width=3.375in]{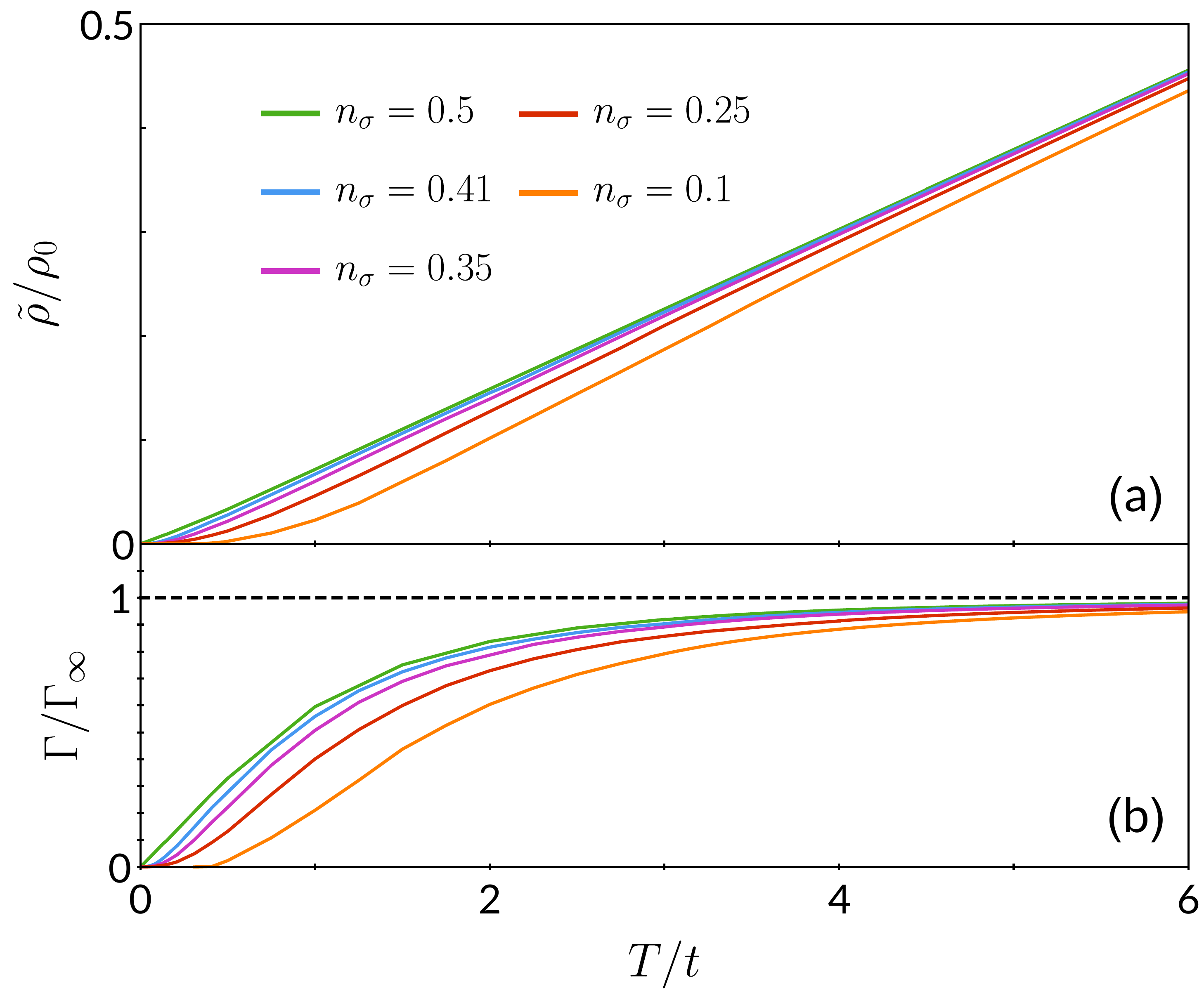}
    \caption{(color online) {(a)} Rescaled resistivity, $\tilde{\rho}=(t/U)^2~\rho$, versus temperature, scaled by $\rho_0=e^2/\hbar$. Data is shown for a variety of densities: $n_\sigma=$ 0.5 (green), 0.41 (blue), 0.35 (purple), 0.25 (red), and 0.1 (orange). The resistivity is a monotonic increasing function of temperature for all densities with a $T$-linear high temperature asymptote, $\tilde{\rho}_\infty/{\rho}_0\approx0.076~T/t$. In the vicinity of half filling, this asymptote describes the resistivity for all temperatures $T\gtrsim t$. {(b)} Scattering rate, $\Gamma$, versus temperature in units of the high-temperature asymptote, $\Gamma_\infty\approx0.609~n_\sigma(1-n_\sigma)~U^2/\hbar t$. At intermediate temperatures, $\Gamma$ changes by about an order of magnitude while the resistivity near half filling shows only small deviations from $T$-linearity. The low-temperature scattering rate has three regimes: $\Gamma\propto T^2$ for $0<|\mu_F|<2t$; $\Gamma\propto T$ for $\mu_F=0$; and $\Gamma\propto e^{-\Delta_U/T}$ for $|\mu_F|>2t$.}
    \label{fig:gammaRes}
\end{figure}

In the weak-coupling limit, the collision integral can be expanded to leading order in powers of the interaction strength, $U/t$. We use only the leading-order term in the expansion, which corresponds to the Born approximation~\cite{coleman}. This results in a resistivity and scattering rate that both scale as $(U/t)^2$. In panel (a) of Figure~\ref{fig:gammaRes} we plot the rescaled resistivity $\tilde\rho(T)=(t/U)^2\rho(T)$ in units of $\rho_0=e^2/\hbar$. Our emphasis is on the intermediate-to-high-temperature regime that is relevant for current cold atom experiments. In panel (b) we plot the quasiparticle scattering rate over the same temperature range. We normalize the scattering rate by its infinite-temperature asymptote, ${\Gamma}_\infty=0.609~n_\sigma(1-n_\sigma)~U^2/\hbar t$, which depends on the particle density. 

The bounded nature of this scatting rate should be contrasted with particles in free space with density $n$ and scattering cross-section $\sigma$.   There $\Gamma=n\sigma \bar v$ continually grows with temperature because the average velocity obeys $\bar v\propto \sqrt{T}$. By contrast, there is a maximum velocity in our single band lattice model which constrains the scattering rate.  

We emphasize that this constraint  is not related to the MIR limit. The MIR limit is defined by the mean free path approaching the lattice spacing, thereby signifying the breakdown of the quasiparticle picture. 
In Ref.~\cite{us} we calculate the infinite-temperature mean free path, finding $l_{\rm MFP}=3.3a(t/U)^2n_\sigma^{-1}(1-n_\sigma)^{-1}\gg a$.

The resistivity in Fig.~\ref{fig:gammaRes} monotonically increases as a function of temperature, a hallmark of metallic behavior. At high temperatures, the resistivity approaches a density-independent $T$-linear asymptote, $\tilde{\rho}_\infty(T)=0.076~(T/t) \rho_0$.  
This behavior
can be understood in terms of the Drude formula $\rho=(m^* \Gamma)/(ne^2)$. 
As the high-temperature scattering rate is a constant, any temperature dependence must come from the 
effective mass. At finite temperature, the relevant weighting of the band curvature is
\begin{equation}
\bigg(\frac{n}{m^*}\bigg)_{\rm eff}=\int \frac{d^2k}{2\pi^2} f_k (\nabla_k^2 \epsilon_k).
\end{equation}
The bottom of the band has positive curvature and makes positive contributions to the inverse effective mass, while the top of the band makes negative contributions. As $T\to\infty$, the distribution function approaches a constant $f_k\to n_\sigma$ and the inverse effective mass vanishes. Leading-order corrections scale as $1/T$, which implies that the effective mass diverges as $m^*\propto T$. As with the upper bound on the scattering rate, this is a generic feature of the single-band Hamiltonian. In cold atom systems,
which are well described by a single band model,
these results imply that a $T$-linear resistivity does not signify a novel strongly-correlated regime in and of itself.


%
%


The high-temperature limiting behavior describes the resistivity down to surprisingly low temperatures, at least an order of magnitude below the bandwidth. This is shown in panel (a) of Fig.~\ref{fig:gammaRes}. For a wide range of fillings, the $T$-linearity continues with the same slope down to temperatures of order $0.1t$.
Comparison with panel (b) shows that the approximate linearity cannot be explained by the high-temperature arguments -- the scattering rate is not constant, changing
by nearly an order of magnitude over the same region. The temperature dependence of the scattering rate, however, is compensated by the temperature dependence of the effective mass. We note that the scattering rate in a system with well-defined quasiparticles is related to the diffusion constant by $D=\frac{1}{2}\langle v^2\rangle/\Gamma$, so this compensation is precisely the same ``conspiracy" observed in Ref.~\cite{Brown379}. Below we discuss a low temperature expansion that sheds light on this.

Resistivity is determined by scattering events that change the total (physical) momentum of the quasiparticles. At high temperatures, the non-linear relationship between the conserved quasimomentum, $\vec{k}$, and the physical momentum, $\hbar\vec{p}/m=\vec\nabla_k\epsilon_k$, means that almost all scattering events contribute to the resistivity. At low temperatures, however, the quasimomentum spectrum can be linearized about the Fermi surface, meaning that quasimomenta are directly proportional to physical momenta. In this case, the only way to change the total physical momentum is for the scattering event to \textit{not} conserve quasimomentum. This can occur via umklapp processes, in which the total incoming and outgoing quasimomenta differ by a reciprocal lattice vector. In the limit $T\to 0$, umklapp processes can only occur when a reciprocal lattice vector is able to connect Fermi surfaces in the extended-zone scheme; for the 2D square lattice with nearest-neighbor hopping, this prohibits umklapp processes when $|\mu_F|>2t$.

An oft-quoted result of Fermi liquid theory is that the low-temperature DC resistivity vanishes as $T^2$. This result follows from a simple phase space argument:
The collision integral involves a sum over the momenta of four particles, $k_j$, that are each constrained to have energies $|\epsilon_j-\mu_F|\lesssim k_BT$. For a given incoming momentum $k_1$ in this range, the number of available values for $k_2$ scale linearly with temperature. Energy and momentum constraints furthermore imply that three of the four degrees of freedom in $k_3$ and $k_4$ are fixed. Thus, the phase space for scattering scales as $T^2$. 

This argument must be modified at half filling due to perfect nesting.  The Fermi surface is a square and 
there are a continuum of umklapp scattering events for fixed $k_1$ and $k_2$, so long as $k_3$ and $k_4$ reside on the half-filled Fermi surface. This acts as an effective constraint at low temperatures, which means that all four degrees of freedom in $k_3$ and $k_4$ are ``fixed" (in that they do not scale with temperature). Hence, the resistivity at half filling vanishes linearly with temperature.


To quantify these low temperature arguments, we expand the Fermi functions about the Fermi surface and integrate the Boltzmann equation~\cite{us} to obtain an expression for the resistivity:
\begin{equation}
    \rho=\frac{\beta}{j^2(\mu_F)}\int_{-8t}^{8t}dE~\frac{(E/2-\mu_F)^2}{\sinh^2(\beta(E/2-\mu_F))}f_T(E)
    \label{eq:lowTRes}
\end{equation}
where $j(\mu_F)$ is the zero-temperature particle number current and $f_T(E)$ is related to the joint density of states for scattering. The current can be expressed in terms of elliptic integrals and  does not contribute to the temperature dependence. At zero temperature the joint density of states is given by $f_{T=0}(E)=\frac{1}{16\pi^4}\sqrt{(4t/E)^2-1}$, which, due to nesting, has a non-integrable divergence at $E= 0$. 
Broadening of the Fermi functions at finite temperature introduces a cutoff which controls this divergence such that $f_T(E=0)\propto 1/T$.  The low temperature behavior for $\mu_F\neq 0$ is seen by replacing the factor $(E/2-\mu_F)^2/\sinh^2(\beta(E/2-\mu_F))$ with a delta function.  

\begin{figure}
    \centering
    \includegraphics[width=3.375in]{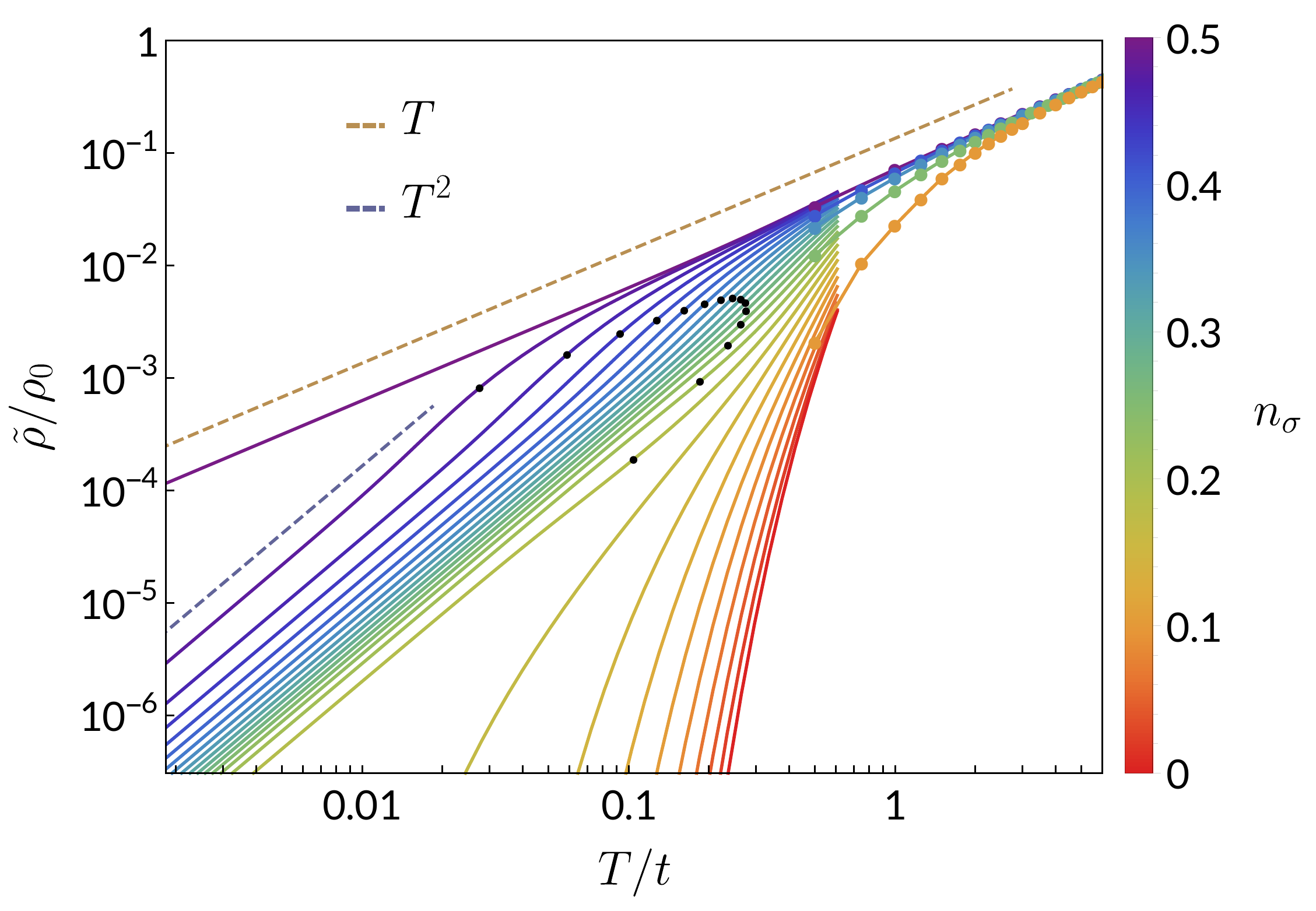}
    \caption{(color online) Rescaled resistivity on a log-log scale for a variety of densities. Solid lines correspond to low-temperature expansion of the resistivity (see Eq~?) and colored dots correspond to intermediate and high-temperature resistivity data (see Fig.~\ref{fig:gammaRes}). The crossover from the high-temperature $T$-linear regime to the three low-temperature regimes is evident. Black dots denote the crossover to $T^2$ behavior for $0<|\mu_F|<2t$. At half filling, the crossover temperature vanishes and $\rho\propto T$ down to $T=0$. The crossover temperature also vanishes as $|\mu_F|\to2t$, above which the Fermi surface is too small to support umklapp processes and the resistivity vanishes exponentially.}
    \label{fig:xTemps}
\end{figure}

In Figure~\ref{fig:xTemps} we plot the resistivity versus temperature on a log-log scale, showing the crossover from low-temperature to high-temperature behavior. Higher-temperature data, shown as colored dots, are calculated using the full Boltzmann equation (cf.  Fig.~\ref{fig:gammaRes}). The solid lines are results from the low-temperature expansion in Eq.~(\ref{eq:lowTRes}). At low temperatures we see three distinct asymptotic behaviors: For $0<|\mu_F|\leq2t$, we recover the expected $\rho\propto T^2$ behavior; for $|\mu_F|>2t$, the resistivity vanishes as $T\exp(-\beta\Delta_U)$, where $\Delta_U=2(|\mu_F|-2t)$ is the ``umklapp gap"~\cite{umklappGap}; and at half-filling, we find $\rho\propto T$.

Even away from half-filling,
mere proximity to the perfectly-nested Fermi surface has a dramatic effect on the  transport properties. In particular, the divergence of the zero-temperature joint-density of states at $E=0$ leads to 
a non-perturbative subleading correction to the $T^2$ resistivity that scales as $T\exp(-2\beta|\mu_F|)$. By equating the leading and subleading terms, we find the scale where the $T^2$ low temperature resistivity crosses over to its $T$-linear high temperature behavior.
This crossover scale is shown in Fig.~\ref{fig:xTemps} by black dots on each curve of fixed density. The crossover temperature vanishes as $|\mu_F|\to 2t$ as the Fermi surface shrinks, but it also vanishes as $\mu_F\to 0$ due to thermal occupation of the nested Fermi surface. The combination of these two effects suppresses the region in which $T^2$ resistivity could be observed
to
$T/t\lesssim 0.1$, almost two orders of magnitude below the bandwidth. This scale is more than a factor of 2 below the lowest temperatures currently accessible in cold atom realizations of the Fermi-Hubbard model~\cite{Brown379,demarco,thywissen}.

We conclude this Letter by emphasizing two points. The first concerns the nature of the MIR limit. The MIR bound on the mean free path only constrains the resistivity when the temperature dependence of the effective mass can be ignored.  As we have shown, this is not the case at high temperatures in models with a finite number of bands: when the temperature is of order the full spectral width, the effective mass falls off as the inverse of the temperature.
While unphysical in the condensed matter context, the regime of near-uniform band filling is common in cold atom experiments.
In these cases, even for systems with a finite scattering rate and well-defined quasiparticles, the resistivity will be unbounded.

The second point concerns the relevance this study to ongoing experiments. Quite rightly, much of the current focus is on the strongly interacting regime, where one can explore the role of correlations.  It is presumptuous, however, to attribute all cases of ``anomalous" transport to strong correlations.  Our results show that even in a weakly interacting gas one would observe $T$-linear resistivity down to temperatures that are a factor of 2 smaller than those achieved in Ref.~\cite{Brown379}. This temperature dependence is simply a band-structure effect and does not give any information about the presence or absence of coherent quasiparticles.

In order to identify a genuine strongly-correlated regime, more direct probes are necessary. One example is the cold atom analog of angle-resolved photoemission spectroscopy, which has been implemented in a quantum gas microscope by the Bakr group~\cite{bakrARPES}. This technique gives direct access to the dispersion and lifetime of quasiparticles and has been valuable in exploring the BCS-BEC crossover~\cite{arpes1,arpes2,arpes3,arpes4,arpes5,arpes6}. Alternatively, the Demarco group has developed a technique to directly access the transport scattering lifetime of the 3D Fermi-Hubbard model~\cite{demarco}. They use a Raman pulse to excite atoms into a finite momentum state and observe the subsequent relaxation. Their technique allows them to probe the MIR limit of the scattering lifetime directly, without reference to the conductivity. The Thywissen group has also studied transport in the 3D model, developing a technique to extract the frequency-dependent optical conductivity $\sigma(\omega)$~\cite{thywissen}. Knowledge of the frequency dependence allows one to explore deviations from Drude-like transport and could be used to distinguish between various theories of the bad metallic phase~\cite{cuprateNS,critpt,holographic}.
They have already studied the DC conductivity in the dilute limit, drawing attention to the umklapp gap.


\begin{acknowledgments}
We thank Debanjan Chowdhury, Joseph Thywissen, and Brian Demarco for helpful conversations.  This material is based upon work supported by the National Science Foundation under Grant No. PHY-1806357 and Grant No. PHY-2110250.
\end{acknowledgments}





\begin{thebibliography}{38}%
\makeatletter
\providecommand \@ifxundefined [1]{%
 \@ifx{#1\undefined}
}%
\providecommand \@ifnum [1]{%
 \ifnum #1\expandafter \@firstoftwo
 \else \expandafter \@secondoftwo
 \fi
}%
\providecommand \@ifx [1]{%
 \ifx #1\expandafter \@firstoftwo
 \else \expandafter \@secondoftwo
 \fi
}%
\providecommand \natexlab [1]{#1}%
\providecommand \enquote  [1]{``#1''}%
\providecommand \bibnamefont  [1]{#1}%
\providecommand \bibfnamefont [1]{#1}%
\providecommand \citenamefont [1]{#1}%
\providecommand \href@noop [0]{\@secondoftwo}%
\providecommand \href [0]{\begingroup \@sanitize@url \@href}%
\providecommand \@href[1]{\@@startlink{#1}\@@href}%
\providecommand \@@href[1]{\endgroup#1\@@endlink}%
\providecommand \@sanitize@url [0]{\catcode `\\12\catcode `\$12\catcode
  `\&12\catcode `\#12\catcode `\^12\catcode `\_12\catcode `\%12\relax}%
\providecommand \@@startlink[1]{}%
\providecommand \@@endlink[0]{}%
\providecommand \url  [0]{\begingroup\@sanitize@url \@url }%
\providecommand \@url [1]{\endgroup\@href {#1}{\urlprefix }}%
\providecommand \urlprefix  [0]{URL }%
\providecommand \Eprint [0]{\href }%
\providecommand \doibase [0]{https://doi.org/}%
\providecommand \selectlanguage [0]{\@gobble}%
\providecommand \bibinfo  [0]{\@secondoftwo}%
\providecommand \bibfield  [0]{\@secondoftwo}%
\providecommand \translation [1]{[#1]}%
\providecommand \BibitemOpen [0]{}%
\providecommand \bibitemStop [0]{}%
\providecommand \bibitemNoStop [0]{.\EOS\space}%
\providecommand \EOS [0]{\spacefactor3000\relax}%
\providecommand \BibitemShut  [1]{\csname bibitem#1\endcsname}%
\let\auto@bib@innerbib\@empty
\bibitem [{\citenamefont {Hussey}(2008)}]{Hussey}%
  \BibitemOpen
  \bibfield  {author} {\bibinfo {author} {\bibfnamefont {N.~E.}\ \bibnamefont
  {Hussey}},\ }\href {https://doi.org/10.1088/0953-8984/20/12/123201}
  {\bibfield  {journal} {\bibinfo  {journal} {Journal of Physics: Condensed
  Matter}\ }\textbf {\bibinfo {volume} {20}},\ \bibinfo {pages} {123201}
  (\bibinfo {year} {2008})}\BibitemShut {NoStop}%
\bibitem [{\citenamefont {Grigera}\ \emph {et~al.}(2001)\citenamefont
  {Grigera}, \citenamefont {Perry}, \citenamefont {Schofield}, \citenamefont
  {Chiao}, \citenamefont {Julian}, \citenamefont {Lonzarich}, \citenamefont
  {Ikeda}, \citenamefont {Maeno}, \citenamefont {Millis},\ and\ \citenamefont
  {Mackenzie}}]{ruthenate}%
  \BibitemOpen
  \bibfield  {author} {\bibinfo {author} {\bibfnamefont {S.~A.}\ \bibnamefont
  {Grigera}}, \bibinfo {author} {\bibfnamefont {R.~S.}\ \bibnamefont {Perry}},
  \bibinfo {author} {\bibfnamefont {A.~J.}\ \bibnamefont {Schofield}}, \bibinfo
  {author} {\bibfnamefont {M.}~\bibnamefont {Chiao}}, \bibinfo {author}
  {\bibfnamefont {S.~R.}\ \bibnamefont {Julian}}, \bibinfo {author}
  {\bibfnamefont {G.~G.}\ \bibnamefont {Lonzarich}}, \bibinfo {author}
  {\bibfnamefont {S.~I.}\ \bibnamefont {Ikeda}}, \bibinfo {author}
  {\bibfnamefont {Y.}~\bibnamefont {Maeno}}, \bibinfo {author} {\bibfnamefont
  {A.~J.}\ \bibnamefont {Millis}},\ and\ \bibinfo {author} {\bibfnamefont
  {A.~P.}\ \bibnamefont {Mackenzie}},\ }\href
  {https://doi.org/10.1126/science.1063539} {\bibfield  {journal} {\bibinfo
  {journal} {Science}\ }\textbf {\bibinfo {volume} {294}},\ \bibinfo {pages}
  {329} (\bibinfo {year} {2001})}\BibitemShut {NoStop}%
\bibitem [{\citenamefont {Doiron-Leyraud}\ \emph {et~al.}(2009)\citenamefont
  {Doiron-Leyraud}, \citenamefont {Auban-Senzier}, \citenamefont {Ren\'e~de
  Cotret}, \citenamefont {Bourbonnais}, \citenamefont {J\'erome}, \citenamefont
  {Bechgaard},\ and\ \citenamefont {Taillefer}}]{pnictide}%
  \BibitemOpen
  \bibfield  {author} {\bibinfo {author} {\bibfnamefont {N.}~\bibnamefont
  {Doiron-Leyraud}}, \bibinfo {author} {\bibfnamefont {P.}~\bibnamefont
  {Auban-Senzier}}, \bibinfo {author} {\bibfnamefont {S.}~\bibnamefont
  {Ren\'e~de Cotret}}, \bibinfo {author} {\bibfnamefont {C.}~\bibnamefont
  {Bourbonnais}}, \bibinfo {author} {\bibfnamefont {D.}~\bibnamefont
  {J\'erome}}, \bibinfo {author} {\bibfnamefont {K.}~\bibnamefont
  {Bechgaard}},\ and\ \bibinfo {author} {\bibfnamefont {L.}~\bibnamefont
  {Taillefer}},\ }\href {https://doi.org/10.1103/PhysRevB.80.214531} {\bibfield
   {journal} {\bibinfo  {journal} {Phys. Rev. B}\ }\textbf {\bibinfo {volume}
  {80}},\ \bibinfo {pages} {214531} (\bibinfo {year} {2009})}\BibitemShut
  {NoStop}%
\bibitem [{\citenamefont {L\"ohneysen}\ \emph {et~al.}(1994)\citenamefont
  {L\"ohneysen}, \citenamefont {Pietrus}, \citenamefont {Portisch},
  \citenamefont {Schlager}, \citenamefont {Schr\"oder}, \citenamefont {Sieck},\
  and\ \citenamefont {Trappmann}}]{heavyFermion}%
  \BibitemOpen
  \bibfield  {author} {\bibinfo {author} {\bibfnamefont {H.~v.}\ \bibnamefont
  {L\"ohneysen}}, \bibinfo {author} {\bibfnamefont {T.}~\bibnamefont
  {Pietrus}}, \bibinfo {author} {\bibfnamefont {G.}~\bibnamefont {Portisch}},
  \bibinfo {author} {\bibfnamefont {H.~G.}\ \bibnamefont {Schlager}}, \bibinfo
  {author} {\bibfnamefont {A.}~\bibnamefont {Schr\"oder}}, \bibinfo {author}
  {\bibfnamefont {M.}~\bibnamefont {Sieck}},\ and\ \bibinfo {author}
  {\bibfnamefont {T.}~\bibnamefont {Trappmann}},\ }\href
  {https://doi.org/10.1103/PhysRevLett.72.3262} {\bibfield  {journal} {\bibinfo
   {journal} {Phys. Rev. Lett.}\ }\textbf {\bibinfo {volume} {72}},\ \bibinfo
  {pages} {3262} (\bibinfo {year} {1994})}\BibitemShut {NoStop}%
\bibitem [{\citenamefont {Legros}\ \emph {et~al.}(2019)\citenamefont {Legros},
  \citenamefont {Benhabib},\ and\ \citenamefont {Tabis}}]{univPlanck}%
  \BibitemOpen
  \bibfield  {author} {\bibinfo {author} {\bibfnamefont {A.}~\bibnamefont
  {Legros}}, \bibinfo {author} {\bibfnamefont {S.}~\bibnamefont {Benhabib}},\
  and\ \bibinfo {author} {\bibfnamefont {W.~e.~a.}\ \bibnamefont {Tabis}},\
  }\href {https://doi.org/10.1038/s41567-018-0334-2} {\bibfield  {journal}
  {\bibinfo  {journal} {Nature Phys}\ }\textbf {\bibinfo {volume} {15}},\
  \bibinfo {pages} {142–147} (\bibinfo {year} {2019})}\BibitemShut {NoStop}%
\bibitem [{\citenamefont {Bruin}\ \emph {et~al.}(2013)\citenamefont {Bruin},
  \citenamefont {Sakai}, \citenamefont {Perry},\ and\ \citenamefont
  {Mackenzie}}]{scattRt}%
  \BibitemOpen
  \bibfield  {author} {\bibinfo {author} {\bibfnamefont {J.~A.~N.}\
  \bibnamefont {Bruin}}, \bibinfo {author} {\bibfnamefont {H.}~\bibnamefont
  {Sakai}}, \bibinfo {author} {\bibfnamefont {R.~S.}\ \bibnamefont {Perry}},\
  and\ \bibinfo {author} {\bibfnamefont {A.~P.}\ \bibnamefont {Mackenzie}},\
  }\href {https://doi.org/10.1126/science.1227612} {\bibfield  {journal}
  {\bibinfo  {journal} {Science}\ }\textbf {\bibinfo {volume} {339}},\ \bibinfo
  {pages} {804} (\bibinfo {year} {2013})}\BibitemShut {NoStop}%
\bibitem [{\citenamefont {Coleman}(2015)}]{coleman}%
  \BibitemOpen
  \bibfield  {author} {\bibinfo {author} {\bibfnamefont {P.}~\bibnamefont
  {Coleman}},\ }\href@noop {} {\emph {\bibinfo {title} {Introduction to
  Many-Body Physics}}}\ (\bibinfo  {publisher} {Cambridge University Press},\
  \bibinfo {year} {2015})\BibitemShut {NoStop}%
\bibitem [{\citenamefont {Mott}(1972)}]{mott}%
  \BibitemOpen
  \bibfield  {author} {\bibinfo {author} {\bibfnamefont {N.~F.}\ \bibnamefont
  {Mott}},\ }\href {https://doi.org/10.1080/14786437208226973} {\bibfield
  {journal} {\bibinfo  {journal} {The Philosophical Magazine: A Journal of
  Theoretical Experimental and Applied Physics}\ }\textbf {\bibinfo {volume}
  {26}},\ \bibinfo {pages} {1015} (\bibinfo {year} {1972})}\BibitemShut
  {NoStop}%
\bibitem [{\citenamefont {Ioffe}\ and\ \citenamefont {Regel}(1960)}]{ioffe}%
  \BibitemOpen
  \bibfield  {author} {\bibinfo {author} {\bibfnamefont {A.~F.}\ \bibnamefont
  {Ioffe}}\ and\ \bibinfo {author} {\bibfnamefont {A.~R.}\ \bibnamefont
  {Regel}},\ }\href@noop {} {\bibfield  {journal} {\bibinfo  {journal} {Prog.
  Semicond.}\ }\textbf {\bibinfo {volume} {4}},\ \bibinfo {pages} {237}
  (\bibinfo {year} {1960})}\BibitemShut {NoStop}%
\bibitem [{\citenamefont {Varma}\ \emph {et~al.}(1989)\citenamefont {Varma},
  \citenamefont {Littlewood}, \citenamefont {Schmitt-Rink}, \citenamefont
  {Abrahams},\ and\ \citenamefont {Ruckenstein}}]{cuprateNS}%
  \BibitemOpen
  \bibfield  {author} {\bibinfo {author} {\bibfnamefont {C.~M.}\ \bibnamefont
  {Varma}}, \bibinfo {author} {\bibfnamefont {P.~B.}\ \bibnamefont
  {Littlewood}}, \bibinfo {author} {\bibfnamefont {S.}~\bibnamefont
  {Schmitt-Rink}}, \bibinfo {author} {\bibfnamefont {E.}~\bibnamefont
  {Abrahams}},\ and\ \bibinfo {author} {\bibfnamefont {A.~E.}\ \bibnamefont
  {Ruckenstein}},\ }\href {https://doi.org/10.1103/PhysRevLett.63.1996}
  {\bibfield  {journal} {\bibinfo  {journal} {Phys. Rev. Lett.}\ }\textbf
  {\bibinfo {volume} {63}},\ \bibinfo {pages} {1996} (\bibinfo {year}
  {1989})}\BibitemShut {NoStop}%
\bibitem [{\citenamefont {Vu\ifmmode \check{c}\else \v{c}\fi{}i\ifmmode
  \check{c}\else \v{c}\fi{}evi\ifmmode~\acute{c}\else \'{c}\fi{}}\ \emph
  {et~al.}(2015)\citenamefont {Vu\ifmmode \check{c}\else \v{c}\fi{}i\ifmmode
  \check{c}\else \v{c}\fi{}evi\ifmmode~\acute{c}\else \'{c}\fi{}},
  \citenamefont {Tanaskovi\ifmmode~\acute{c}\else \'{c}\fi{}}, \citenamefont
  {Rozenberg},\ and\ \citenamefont {Dobrosavljevi\ifmmode~\acute{c}\else
  \'{c}\fi{}}}]{critpt}%
  \BibitemOpen
  \bibfield  {author} {\bibinfo {author} {\bibfnamefont {J.}~\bibnamefont
  {Vu\ifmmode \check{c}\else \v{c}\fi{}i\ifmmode \check{c}\else
  \v{c}\fi{}evi\ifmmode~\acute{c}\else \'{c}\fi{}}}, \bibinfo {author}
  {\bibfnamefont {D.}~\bibnamefont {Tanaskovi\ifmmode~\acute{c}\else
  \'{c}\fi{}}}, \bibinfo {author} {\bibfnamefont {M.~J.}\ \bibnamefont
  {Rozenberg}},\ and\ \bibinfo {author} {\bibfnamefont {V.}~\bibnamefont
  {Dobrosavljevi\ifmmode~\acute{c}\else \'{c}\fi{}}},\ }\href
  {https://doi.org/10.1103/PhysRevLett.114.246402} {\bibfield  {journal}
  {\bibinfo  {journal} {Phys. Rev. Lett.}\ }\textbf {\bibinfo {volume} {114}},\
  \bibinfo {pages} {246402} (\bibinfo {year} {2015})}\BibitemShut {NoStop}%
\bibitem [{\citenamefont {Hartnoll}\ \emph {et~al.}(2018)\citenamefont
  {Hartnoll}, \citenamefont {Lucas},\ and\ \citenamefont
  {Sachdev}}]{holographic}%
  \BibitemOpen
  \bibfield  {author} {\bibinfo {author} {\bibfnamefont {S.}~\bibnamefont
  {Hartnoll}}, \bibinfo {author} {\bibfnamefont {A.}~\bibnamefont {Lucas}},\
  and\ \bibinfo {author} {\bibfnamefont {S.}~\bibnamefont {Sachdev}},\
  }\href@noop {} {\emph {\bibinfo {title} {Holographic Quantum Matter}}}\
  (\bibinfo  {publisher} {MIT Press},\ \bibinfo {year} {2018})\BibitemShut
  {NoStop}%
\bibitem [{\citenamefont {Perepelitsky}\ \emph {et~al.}(2016)\citenamefont
  {Perepelitsky}, \citenamefont {Galatas}, \citenamefont {Mravlje},
  \citenamefont {\ifmmode~\check{Z}\else \v{Z}\fi{}itko}, \citenamefont
  {Khatami}, \citenamefont {Shastry},\ and\ \citenamefont
  {Georges}}]{highTPerspective}%
  \BibitemOpen
  \bibfield  {author} {\bibinfo {author} {\bibfnamefont {E.}~\bibnamefont
  {Perepelitsky}}, \bibinfo {author} {\bibfnamefont {A.}~\bibnamefont
  {Galatas}}, \bibinfo {author} {\bibfnamefont {J.}~\bibnamefont {Mravlje}},
  \bibinfo {author} {\bibfnamefont {R.}~\bibnamefont {\ifmmode~\check{Z}\else
  \v{Z}\fi{}itko}}, \bibinfo {author} {\bibfnamefont {E.}~\bibnamefont
  {Khatami}}, \bibinfo {author} {\bibfnamefont {B.~S.}\ \bibnamefont
  {Shastry}},\ and\ \bibinfo {author} {\bibfnamefont {A.}~\bibnamefont
  {Georges}},\ }\href {https://doi.org/10.1103/PhysRevB.94.235115} {\bibfield
  {journal} {\bibinfo  {journal} {Phys. Rev. B}\ }\textbf {\bibinfo {volume}
  {94}},\ \bibinfo {pages} {235115} (\bibinfo {year} {2016})}\BibitemShut
  {NoStop}%
\bibitem [{\citenamefont {Mousatov}\ \emph {et~al.}(2019)\citenamefont
  {Mousatov}, \citenamefont {Esterlis},\ and\ \citenamefont
  {Hartnoll}}]{hartnollBadMetal}%
  \BibitemOpen
  \bibfield  {author} {\bibinfo {author} {\bibfnamefont {C.~H.}\ \bibnamefont
  {Mousatov}}, \bibinfo {author} {\bibfnamefont {I.}~\bibnamefont {Esterlis}},\
  and\ \bibinfo {author} {\bibfnamefont {S.~A.}\ \bibnamefont {Hartnoll}},\
  }\href {https://doi.org/10.1103/PhysRevLett.122.186601} {\bibfield  {journal}
  {\bibinfo  {journal} {Phys. Rev. Lett.}\ }\textbf {\bibinfo {volume} {122}},\
  \bibinfo {pages} {186601} (\bibinfo {year} {2019})}\BibitemShut {NoStop}%
\bibitem [{\citenamefont {Vu\ifmmode \check{c}\else \v{c}\fi{}i\ifmmode
  \check{c}\else \v{c}\fi{}evi\ifmmode~\acute{c}\else \'{c}\fi{}}\ \emph
  {et~al.}(2019)\citenamefont {Vu\ifmmode \check{c}\else \v{c}\fi{}i\ifmmode
  \check{c}\else \v{c}\fi{}evi\ifmmode~\acute{c}\else \'{c}\fi{}},
  \citenamefont {Kokalj}, \citenamefont {\ifmmode~\check{Z}\else
  \v{Z}\fi{}itko}, \citenamefont {Wentzell}, \citenamefont
  {Tanaskovi\ifmmode~\acute{c}\else \'{c}\fi{}},\ and\ \citenamefont
  {Mravlje}}]{numericalHubbardVertex}%
  \BibitemOpen
  \bibfield  {author} {\bibinfo {author} {\bibfnamefont {J.}~\bibnamefont
  {Vu\ifmmode \check{c}\else \v{c}\fi{}i\ifmmode \check{c}\else
  \v{c}\fi{}evi\ifmmode~\acute{c}\else \'{c}\fi{}}}, \bibinfo {author}
  {\bibfnamefont {J.}~\bibnamefont {Kokalj}}, \bibinfo {author} {\bibfnamefont
  {R.}~\bibnamefont {\ifmmode~\check{Z}\else \v{Z}\fi{}itko}}, \bibinfo
  {author} {\bibfnamefont {N.}~\bibnamefont {Wentzell}}, \bibinfo {author}
  {\bibfnamefont {D.}~\bibnamefont {Tanaskovi\ifmmode~\acute{c}\else
  \'{c}\fi{}}},\ and\ \bibinfo {author} {\bibfnamefont {J.}~\bibnamefont
  {Mravlje}},\ }\href {https://doi.org/10.1103/PhysRevLett.123.036601}
  {\bibfield  {journal} {\bibinfo  {journal} {Phys. Rev. Lett.}\ }\textbf
  {\bibinfo {volume} {123}},\ \bibinfo {pages} {036601} (\bibinfo {year}
  {2019})}\BibitemShut {NoStop}%
\bibitem [{\citenamefont {Huang}\ \emph {et~al.}(2019)\citenamefont {Huang},
  \citenamefont {Sheppard}, \citenamefont {Moritz},\ and\ \citenamefont
  {Devereaux}}]{Huang}%
  \BibitemOpen
  \bibfield  {author} {\bibinfo {author} {\bibfnamefont {E.~W.}\ \bibnamefont
  {Huang}}, \bibinfo {author} {\bibfnamefont {R.}~\bibnamefont {Sheppard}},
  \bibinfo {author} {\bibfnamefont {B.}~\bibnamefont {Moritz}},\ and\ \bibinfo
  {author} {\bibfnamefont {T.~P.}\ \bibnamefont {Devereaux}},\ }\href
  {https://doi.org/10.1126/science.aau7063} {\bibfield  {journal} {\bibinfo
  {journal} {Science}\ }\textbf {\bibinfo {volume} {366}},\ \bibinfo {pages}
  {987} (\bibinfo {year} {2019})}\BibitemShut {NoStop}%
\bibitem [{\citenamefont {Cha}\ \emph {et~al.}(2020)\citenamefont {Cha},
  \citenamefont {Patel}, \citenamefont {Gull},\ and\ \citenamefont
  {Kim}}]{slopeInvariantTLinear}%
  \BibitemOpen
  \bibfield  {author} {\bibinfo {author} {\bibfnamefont {P.}~\bibnamefont
  {Cha}}, \bibinfo {author} {\bibfnamefont {A.~A.}\ \bibnamefont {Patel}},
  \bibinfo {author} {\bibfnamefont {E.}~\bibnamefont {Gull}},\ and\ \bibinfo
  {author} {\bibfnamefont {E.-A.}\ \bibnamefont {Kim}},\ }\href
  {https://doi.org/10.1103/PhysRevResearch.2.033434} {\bibfield  {journal}
  {\bibinfo  {journal} {Phys. Rev. Research}\ }\textbf {\bibinfo {volume}
  {2}},\ \bibinfo {pages} {033434} (\bibinfo {year} {2020})}\BibitemShut
  {NoStop}%
\bibitem [{\citenamefont {Bloch}\ \emph {et~al.}(2008)\citenamefont {Bloch},
  \citenamefont {Dalibard},\ and\ \citenamefont {Zwerger}}]{expReview1}%
  \BibitemOpen
  \bibfield  {author} {\bibinfo {author} {\bibfnamefont {I.}~\bibnamefont
  {Bloch}}, \bibinfo {author} {\bibfnamefont {J.}~\bibnamefont {Dalibard}},\
  and\ \bibinfo {author} {\bibfnamefont {W.}~\bibnamefont {Zwerger}},\ }\href
  {https://doi.org/10.1103/RevModPhys.80.885} {\bibfield  {journal} {\bibinfo
  {journal} {Rev. Mod. Phys.}\ }\textbf {\bibinfo {volume} {80}},\ \bibinfo
  {pages} {885} (\bibinfo {year} {2008})}\BibitemShut {NoStop}%
\bibitem [{\citenamefont {Chin}\ \emph {et~al.}(2010)\citenamefont {Chin},
  \citenamefont {Grimm}, \citenamefont {Julienne},\ and\ \citenamefont
  {Tiesinga}}]{expReview2}%
  \BibitemOpen
  \bibfield  {author} {\bibinfo {author} {\bibfnamefont {C.}~\bibnamefont
  {Chin}}, \bibinfo {author} {\bibfnamefont {R.}~\bibnamefont {Grimm}},
  \bibinfo {author} {\bibfnamefont {P.}~\bibnamefont {Julienne}},\ and\
  \bibinfo {author} {\bibfnamefont {E.}~\bibnamefont {Tiesinga}},\ }\href
  {https://doi.org/10.1103/RevModPhys.82.1225} {\bibfield  {journal} {\bibinfo
  {journal} {Rev. Mod. Phys.}\ }\textbf {\bibinfo {volume} {82}},\ \bibinfo
  {pages} {1225} (\bibinfo {year} {2010})}\BibitemShut {NoStop}%
\bibitem [{\citenamefont {Brown}\ \emph {et~al.}(2019)\citenamefont {Brown},
  \citenamefont {Mitra}, \citenamefont {Guardado-Sanchez}, \citenamefont
  {Nourafkan}, \citenamefont {Reymbaut}, \citenamefont {H{\'e}bert},
  \citenamefont {Bergeron}, \citenamefont {Tremblay}, \citenamefont {Kokalj},
  \citenamefont {Huse}, \citenamefont {Schau{\ss}},\ and\ \citenamefont
  {Bakr}}]{Brown379}%
  \BibitemOpen
  \bibfield  {author} {\bibinfo {author} {\bibfnamefont {P.~T.}\ \bibnamefont
  {Brown}}, \bibinfo {author} {\bibfnamefont {D.}~\bibnamefont {Mitra}},
  \bibinfo {author} {\bibfnamefont {E.}~\bibnamefont {Guardado-Sanchez}},
  \bibinfo {author} {\bibfnamefont {R.}~\bibnamefont {Nourafkan}}, \bibinfo
  {author} {\bibfnamefont {A.}~\bibnamefont {Reymbaut}}, \bibinfo {author}
  {\bibfnamefont {C.-D.}\ \bibnamefont {H{\'e}bert}}, \bibinfo {author}
  {\bibfnamefont {S.}~\bibnamefont {Bergeron}}, \bibinfo {author}
  {\bibfnamefont {A.-M.~S.}\ \bibnamefont {Tremblay}}, \bibinfo {author}
  {\bibfnamefont {J.}~\bibnamefont {Kokalj}}, \bibinfo {author} {\bibfnamefont
  {D.~A.}\ \bibnamefont {Huse}}, \bibinfo {author} {\bibfnamefont
  {P.}~\bibnamefont {Schau{\ss}}},\ and\ \bibinfo {author} {\bibfnamefont
  {W.~S.}\ \bibnamefont {Bakr}},\ }\href
  {https://doi.org/10.1126/science.aat4134} {\bibfield  {journal} {\bibinfo
  {journal} {Science}\ }\textbf {\bibinfo {volume} {363}},\ \bibinfo {pages}
  {379} (\bibinfo {year} {2019})}\BibitemShut {NoStop}%
\bibitem [{\citenamefont {Xu}\ \emph {et~al.}(2019)\citenamefont {Xu},
  \citenamefont {McGehee}, \citenamefont {Morong},\ and\ \citenamefont
  {DeMarco}}]{demarco}%
  \BibitemOpen
  \bibfield  {author} {\bibinfo {author} {\bibfnamefont {W.}~\bibnamefont
  {Xu}}, \bibinfo {author} {\bibfnamefont {W.}~\bibnamefont {McGehee}},
  \bibinfo {author} {\bibfnamefont {W.}~\bibnamefont {Morong}},\ and\ \bibinfo
  {author} {\bibfnamefont {B.}~\bibnamefont {DeMarco}},\ }\href
  {https://doi.org/10.1038/s41467-019-09526-x} {\bibfield  {journal} {\bibinfo
  {journal} {Nature communications}\ }\textbf {\bibinfo {volume} {10}},\
  \bibinfo {pages} {1588} (\bibinfo {year} {2019})}\BibitemShut {NoStop}%
\bibitem [{\citenamefont {Anderson}\ \emph {et~al.}(2019)\citenamefont
  {Anderson}, \citenamefont {Wang}, \citenamefont {Xu}, \citenamefont {Venu},
  \citenamefont {Trotzky}, \citenamefont {Chevy},\ and\ \citenamefont
  {Thywissen}}]{thywissen}%
  \BibitemOpen
  \bibfield  {author} {\bibinfo {author} {\bibfnamefont {R.}~\bibnamefont
  {Anderson}}, \bibinfo {author} {\bibfnamefont {F.}~\bibnamefont {Wang}},
  \bibinfo {author} {\bibfnamefont {P.}~\bibnamefont {Xu}}, \bibinfo {author}
  {\bibfnamefont {V.}~\bibnamefont {Venu}}, \bibinfo {author} {\bibfnamefont
  {S.}~\bibnamefont {Trotzky}}, \bibinfo {author} {\bibfnamefont
  {F.}~\bibnamefont {Chevy}},\ and\ \bibinfo {author} {\bibfnamefont {J.~H.}\
  \bibnamefont {Thywissen}},\ }\href
  {https://doi.org/10.1103/PhysRevLett.122.153602} {\bibfield  {journal}
  {\bibinfo  {journal} {Phys. Rev. Lett.}\ }\textbf {\bibinfo {volume} {122}},\
  \bibinfo {pages} {153602} (\bibinfo {year} {2019})}\BibitemShut {NoStop}%
\bibitem [{\citenamefont {Anderson}(1987)}]{ANDERSON1196}%
  \BibitemOpen
  \bibfield  {author} {\bibinfo {author} {\bibfnamefont {P.~W.}\ \bibnamefont
  {Anderson}},\ }\href {https://doi.org/10.1126/science.235.4793.1196}
  {\bibfield  {journal} {\bibinfo  {journal} {Science}\ }\textbf {\bibinfo
  {volume} {235}},\ \bibinfo {pages} {1196} (\bibinfo {year}
  {1987})}\BibitemShut {NoStop}%
\bibitem [{\citenamefont {Arovas}\ \emph {et~al.}(2021)\citenamefont {Arovas},
  \citenamefont {Berg}, \citenamefont {Kivelson},\ and\ \citenamefont
  {Raghu}}]{arovas2021hubbard}%
  \BibitemOpen
  \bibfield  {author} {\bibinfo {author} {\bibfnamefont {D.~P.}\ \bibnamefont
  {Arovas}}, \bibinfo {author} {\bibfnamefont {E.}~\bibnamefont {Berg}},
  \bibinfo {author} {\bibfnamefont {S.}~\bibnamefont {Kivelson}},\ and\
  \bibinfo {author} {\bibfnamefont {S.}~\bibnamefont {Raghu}},\ }\href@noop {}
  {\bibinfo {title} {The hubbard model}} (\bibinfo {year} {2021}),\ \Eprint
  {https://arxiv.org/abs/2103.12097} {arXiv:2103.12097 [cond-mat.str-el]}
  \BibitemShut {NoStop}%
\bibitem [{\citenamefont {Qin}\ \emph {et~al.}(2021)\citenamefont {Qin},
  \citenamefont {Schäfer}, \citenamefont {Andergassen}, \citenamefont
  {Corboz},\ and\ \citenamefont {Gull}}]{computationalPerspective}%
  \BibitemOpen
  \bibfield  {author} {\bibinfo {author} {\bibfnamefont {M.}~\bibnamefont
  {Qin}}, \bibinfo {author} {\bibfnamefont {T.}~\bibnamefont {Schäfer}},
  \bibinfo {author} {\bibfnamefont {S.}~\bibnamefont {Andergassen}}, \bibinfo
  {author} {\bibfnamefont {P.}~\bibnamefont {Corboz}},\ and\ \bibinfo {author}
  {\bibfnamefont {E.}~\bibnamefont {Gull}},\ }\href@noop {} {\bibinfo {title}
  {The hubbard model: A computational perspective}} (\bibinfo {year} {2021}),\
  \Eprint {https://arxiv.org/abs/2104.00064} {arXiv:2104.00064
  [cond-mat.str-el]} \BibitemShut {NoStop}%
\bibitem [{\citenamefont {Ziman}(1960)}]{Ziman}%
  \BibitemOpen
  \bibfield  {author} {\bibinfo {author} {\bibfnamefont {J.~M.}\ \bibnamefont
  {Ziman}},\ }\href@noop {} {\emph {\bibinfo {title} {Electrons and Phonons:
  The Theory of Transport Phenomena in Solids}}}\ (\bibinfo  {publisher}
  {Oxford University Press},\ \bibinfo {year} {1960})\BibitemShut {NoStop}%
\bibitem [{\citenamefont {Ziman}(1956)}]{Ziman2}%
  \BibitemOpen
  \bibfield  {author} {\bibinfo {author} {\bibfnamefont {J.~M.}\ \bibnamefont
  {Ziman}},\ }\href {https://doi.org/10.1139/p56-139} {\bibfield  {journal}
  {\bibinfo  {journal} {Canadian Journal of Physics}\ }\textbf {\bibinfo
  {volume} {34}},\ \bibinfo {pages} {1256} (\bibinfo {year} {1956})},\ \Eprint
  {https://arxiv.org/abs/https://doi.org/10.1139/p56-139}
  {https://doi.org/10.1139/p56-139} \BibitemShut {NoStop}%
\bibitem [{\citenamefont {Kiely}\ and\ \citenamefont {Mueller}(2021)}]{us}%
  \BibitemOpen
  \bibfield  {author} {\bibinfo {author} {\bibfnamefont {T.~G.}\ \bibnamefont
  {Kiely}}\ and\ \bibinfo {author} {\bibfnamefont {E.~J.}\ \bibnamefont
  {Mueller}},\ }\href@noop {} {\bibinfo {title} {Transport in the 2d
  fermi-hubbard model: Lessons from weak coupling}} (\bibinfo {year} {2021}),\
  \Eprint {https://arxiv.org/abs/2106.04479} {arXiv:2106.04479
  [cond-mat.quant-gas]} \BibitemShut {NoStop}%
\bibitem [{\citenamefont {Onsager}(1931{\natexlab{a}})}]{onsager1}%
  \BibitemOpen
  \bibfield  {author} {\bibinfo {author} {\bibfnamefont {L.}~\bibnamefont
  {Onsager}},\ }\href {https://doi.org/10.1103/PhysRev.37.405} {\bibfield
  {journal} {\bibinfo  {journal} {Phys. Rev.}\ }\textbf {\bibinfo {volume}
  {37}},\ \bibinfo {pages} {405} (\bibinfo {year}
  {1931}{\natexlab{a}})}\BibitemShut {NoStop}%
\bibitem [{\citenamefont {Onsager}(1931{\natexlab{b}})}]{onsager2}%
  \BibitemOpen
  \bibfield  {author} {\bibinfo {author} {\bibfnamefont {L.}~\bibnamefont
  {Onsager}},\ }\href {https://doi.org/10.1103/PhysRev.38.2265} {\bibfield
  {journal} {\bibinfo  {journal} {Phys. Rev.}\ }\textbf {\bibinfo {volume}
  {38}},\ \bibinfo {pages} {2265} (\bibinfo {year}
  {1931}{\natexlab{b}})}\BibitemShut {NoStop}%
\bibitem [{\citenamefont {Rosch}(2006)}]{umklappGap}%
  \BibitemOpen
  \bibfield  {author} {\bibinfo {author} {\bibfnamefont {A.}~\bibnamefont
  {Rosch}},\ }\href {https://doi.org/https://doi.org/10.1002/andp.200510203}
  {\bibfield  {journal} {\bibinfo  {journal} {Annalen der Physik}\ }\textbf
  {\bibinfo {volume} {15}},\ \bibinfo {pages} {526} (\bibinfo {year}
  {2006})}\BibitemShut {NoStop}%
\bibitem [{\citenamefont {Brown}\ \emph {et~al.}(2020)\citenamefont {Brown},
  \citenamefont {Guardado-Sanchez}, \citenamefont {Spar}, \citenamefont
  {Huang}, \citenamefont {Devereaux},\ and\ \citenamefont {Bakr}}]{bakrARPES}%
  \BibitemOpen
  \bibfield  {author} {\bibinfo {author} {\bibfnamefont {P.}~\bibnamefont
  {Brown}}, \bibinfo {author} {\bibfnamefont {E.}~\bibnamefont
  {Guardado-Sanchez}}, \bibinfo {author} {\bibfnamefont {B.}~\bibnamefont
  {Spar}}, \bibinfo {author} {\bibfnamefont {E.}~\bibnamefont {Huang}},
  \bibinfo {author} {\bibfnamefont {T.}~\bibnamefont {Devereaux}},\ and\
  \bibinfo {author} {\bibfnamefont {W.}~\bibnamefont {Bakr}},\ }\href
  {https://doi.org/10.1038/s41567-019-0696-0} {\bibfield  {journal} {\bibinfo
  {journal} {Nat. Phys.}\ }\textbf {\bibinfo {volume} {16}},\ \bibinfo {pages}
  {26–31} (\bibinfo {year} {2020})}\BibitemShut {NoStop}%
\bibitem [{\citenamefont {Dao}\ \emph {et~al.}(2007)\citenamefont {Dao},
  \citenamefont {Georges}, \citenamefont {Dalibard}, \citenamefont {Salomon},\
  and\ \citenamefont {Carusotto}}]{arpes1}%
  \BibitemOpen
  \bibfield  {author} {\bibinfo {author} {\bibfnamefont {T.-L.}\ \bibnamefont
  {Dao}}, \bibinfo {author} {\bibfnamefont {A.}~\bibnamefont {Georges}},
  \bibinfo {author} {\bibfnamefont {J.}~\bibnamefont {Dalibard}}, \bibinfo
  {author} {\bibfnamefont {C.}~\bibnamefont {Salomon}},\ and\ \bibinfo {author}
  {\bibfnamefont {I.}~\bibnamefont {Carusotto}},\ }\href
  {https://doi.org/10.1103/PhysRevLett.98.240402} {\bibfield  {journal}
  {\bibinfo  {journal} {Phys. Rev. Lett.}\ }\textbf {\bibinfo {volume} {98}},\
  \bibinfo {pages} {240402} (\bibinfo {year} {2007})}\BibitemShut {NoStop}%
\bibitem [{\citenamefont {Stewart}\ \emph {et~al.}(2008)\citenamefont
  {Stewart}, \citenamefont {Gaebler},\ and\ \citenamefont {Jin}}]{arpes2}%
  \BibitemOpen
  \bibfield  {author} {\bibinfo {author} {\bibfnamefont {J.}~\bibnamefont
  {Stewart}}, \bibinfo {author} {\bibfnamefont {J.}~\bibnamefont {Gaebler}},\
  and\ \bibinfo {author} {\bibfnamefont {D.}~\bibnamefont {Jin}},\ }\href
  {https://doi.org/10.1038/nature07172} {\bibfield  {journal} {\bibinfo
  {journal} {Nature}\ }\textbf {\bibinfo {volume} {454}},\ \bibinfo {pages}
  {744–747} (\bibinfo {year} {2008})}\BibitemShut {NoStop}%
\bibitem [{\citenamefont {Gaebler}\ \emph {et~al.}(2010)\citenamefont
  {Gaebler}, \citenamefont {Stewart}, \citenamefont {Drake}, \citenamefont
  {Jin}, \citenamefont {Perali}, \citenamefont {Pieri},\ and\ \citenamefont
  {Strinati}}]{arpes3}%
  \BibitemOpen
  \bibfield  {author} {\bibinfo {author} {\bibfnamefont {J.}~\bibnamefont
  {Gaebler}}, \bibinfo {author} {\bibfnamefont {J.}~\bibnamefont {Stewart}},
  \bibinfo {author} {\bibfnamefont {T.}~\bibnamefont {Drake}}, \bibinfo
  {author} {\bibfnamefont {D.}~\bibnamefont {Jin}}, \bibinfo {author}
  {\bibfnamefont {A.}~\bibnamefont {Perali}}, \bibinfo {author} {\bibfnamefont
  {P.}~\bibnamefont {Pieri}},\ and\ \bibinfo {author} {\bibfnamefont
  {G.}~\bibnamefont {Strinati}},\ }\href {https://doi.org/10.1038/nphys1709}
  {\bibfield  {journal} {\bibinfo  {journal} {Nature Phys}\ }\textbf {\bibinfo
  {volume} {6}},\ \bibinfo {pages} {569–573} (\bibinfo {year}
  {2010})}\BibitemShut {NoStop}%
\bibitem [{\citenamefont {Feld}\ \emph {et~al.}(2011)\citenamefont {Feld},
  \citenamefont {Fröhlich}, \citenamefont {Vogt}, \citenamefont
  {Koschorreck},\ and\ \citenamefont {Kohl}}]{arpes4}%
  \BibitemOpen
  \bibfield  {author} {\bibinfo {author} {\bibfnamefont {M.}~\bibnamefont
  {Feld}}, \bibinfo {author} {\bibfnamefont {B.}~\bibnamefont {Fröhlich}},
  \bibinfo {author} {\bibfnamefont {E.}~\bibnamefont {Vogt}}, \bibinfo {author}
  {\bibfnamefont {M.}~\bibnamefont {Koschorreck}},\ and\ \bibinfo {author}
  {\bibfnamefont {M.}~\bibnamefont {Kohl}},\ }\href
  {https://doi.org/10.1038/nature10627} {\bibfield  {journal} {\bibinfo
  {journal} {Nature}\ }\textbf {\bibinfo {volume} {480}},\ \bibinfo {pages}
  {75–78} (\bibinfo {year} {2011})}\BibitemShut {NoStop}%
\bibitem [{\citenamefont {Fr\"ohlich}\ \emph {et~al.}(2012)\citenamefont
  {Fr\"ohlich}, \citenamefont {Feld}, \citenamefont {Vogt}, \citenamefont
  {Koschorreck}, \citenamefont {K\"ohl}, \citenamefont {Berthod},\ and\
  \citenamefont {Giamarchi}}]{arpes5}%
  \BibitemOpen
  \bibfield  {author} {\bibinfo {author} {\bibfnamefont {B.}~\bibnamefont
  {Fr\"ohlich}}, \bibinfo {author} {\bibfnamefont {M.}~\bibnamefont {Feld}},
  \bibinfo {author} {\bibfnamefont {E.}~\bibnamefont {Vogt}}, \bibinfo {author}
  {\bibfnamefont {M.}~\bibnamefont {Koschorreck}}, \bibinfo {author}
  {\bibfnamefont {M.}~\bibnamefont {K\"ohl}}, \bibinfo {author} {\bibfnamefont
  {C.}~\bibnamefont {Berthod}},\ and\ \bibinfo {author} {\bibfnamefont
  {T.}~\bibnamefont {Giamarchi}},\ }\href
  {https://doi.org/10.1103/PhysRevLett.109.130403} {\bibfield  {journal}
  {\bibinfo  {journal} {Phys. Rev. Lett.}\ }\textbf {\bibinfo {volume} {109}},\
  \bibinfo {pages} {130403} (\bibinfo {year} {2012})}\BibitemShut {NoStop}%
\bibitem [{\citenamefont {Sagi}\ \emph {et~al.}(2015)\citenamefont {Sagi},
  \citenamefont {Drake}, \citenamefont {Paudel}, \citenamefont {Chapurin},\
  and\ \citenamefont {Jin}}]{arpes6}%
  \BibitemOpen
  \bibfield  {author} {\bibinfo {author} {\bibfnamefont {Y.}~\bibnamefont
  {Sagi}}, \bibinfo {author} {\bibfnamefont {T.~E.}\ \bibnamefont {Drake}},
  \bibinfo {author} {\bibfnamefont {R.}~\bibnamefont {Paudel}}, \bibinfo
  {author} {\bibfnamefont {R.}~\bibnamefont {Chapurin}},\ and\ \bibinfo
  {author} {\bibfnamefont {D.~S.}\ \bibnamefont {Jin}},\ }\href
  {https://doi.org/10.1103/PhysRevLett.114.075301} {\bibfield  {journal}
  {\bibinfo  {journal} {Phys. Rev. Lett.}\ }\textbf {\bibinfo {volume} {114}},\
  \bibinfo {pages} {075301} (\bibinfo {year} {2015})}\BibitemShut {NoStop}%
\end{thebibliography}

\providecommand{\noopsort}[1]{}\providecommand{\singleletter}[1]{#1}%

\end{document}